\documentstyle[aps,eqsecnum,epsf]{revtex}

\begin{document}
\thispagestyle{empty}
\begin{flushright} JLAB-THY-98-31 \end{flushright}
\vspace{7mm}
\begin{center}
{\Large{\bf The $P_{33}(1232)$ resonance
contribution into the amplitudes
$M_{1+}^{3/2},E_{1+}^{3/2},S_{1+}^{3/2}$
from an analysis of the
$p(e,e'p)\pi^0$ data
at $Q^2~=$ 2.8, 3.2, and 4 $(GeV/c)^2$
within dispersion relation approach}}\\
\vspace{7mm}
{\large I.G.Aznauryan}\\
\vspace{7mm}
{\em Yerevan Physics Institute,
Alikhanian Brothers St.2, Yerevan, 375036 Armenia}\\
{(e-mail addresses: aznaur@jerewan1.yerphi.am, aznaury@cebaf.gov)}\\
\vspace{7mm}
{\large S.G.Stepanyan}\\
\vspace{7mm}
{\em Jefferson Lab,
12000 Jefferson Avenue, Newport News, VA 23606, USA,}\\
{\em Yerevan Physics Institute,
Alikhanian Brothers St.2, Yerevan, 375036 Armenia}\\
{(e-mail address: stepanya@cebaf.gov)}\\

\end{center}

\begin{abstract}
Within the fixed-t dispersion relation approach
we have analysed  the TJNAF and DESY data on the exclusive $p(e,e'p)\pi^0$
reaction in order to find the $P_{33}(1232)$
resonance contribution into the multipole amplitudes
$M_{1+}^{3/2},E_{1+}^{3/2},S_{1+}^{3/2}$.
As an input for the resonance and nonresonance
contributions into these amplitudes
the earlier obtained
solutions of the integral equations
which follow from dispersion relations are used.
The obtained values of the ratio $E2/M1$
for the  $\gamma^* N \rightarrow P_{33}(1232)$ transition
are: $0.039\pm 0.029,~0.121\pm 0.032,~0.04\pm 0.031$
for  $Q^2~=$ 2.8, 3.2, and 4 $(GeV/c)^2$, respectively.
The comparison with the data at low  $Q^2$
shows that there is no evidence for the presence of
the visible
pQCD contribution into the transition
$\gamma N \rightarrow P_{33}(1232)$
at $Q^2=3-4~GeV^2$.
The ratio $S_{1+}^{3/2}/M_{1+}^{3/2}$ for the 
resonance parts of multipoles is:
$-0.049\pm 0.029,~-0.099\pm 0.041,~-0.085\pm 0.021$
for  $Q^2~=$ 2.8, 3.2, and 4 $(GeV/c)^2$, respectively.
Our results for the transverse form factor $G_T(Q^2)$ of the 
$\gamma^* N \rightarrow P_{33}(1232)$ transition
are lower than the values obtained
from the inclusive data.
With increasing $Q^2$, $Q^4G_T(Q^2)$ decreases, 
so there is no evidence for the presence
of the pQCD contribution here too.
\end{abstract}

\vspace{3mm}
PACS number(s): 11.55.Fv, 11.80.Et, 13.60.Le, 25.20.Lj, 25.30Rw 
\vspace{3mm}

\section{Introduction}
It is known that the information on the $Q^2$ evolution
of the $\gamma^* N \rightarrow P_{33}(1232)$
transition form factors
may play an important role
in the investigation of the energetic scale of the transition
to the perturbative region of QCD.
Especially
important is the information on the
ratio $E2/M1$ which being close to 0
at small $Q^2$ should go to 1
in the pQCD asymptotics.
Experimental data on the 
cross sections of the exclusive reaction
$p(e,e'p)\pi^0$
obtained recently at TJNAF at
$Q^2~=$ 2.8 and 4 $(GeV/c)^2$ \cite{1}
and more earlier DESY data at
$Q^2~=~3.2~(GeV/c)^2$ \cite{2}
can be useful for understanding of the place
of the region of $Q^2~=~3-4~GeV^2$
in the transition to the pQCD regime.
These data will be analysed in the present 
paper in order to extract an information
on the $\gamma^* N \rightarrow P_{33}(1232)$
transition in this region of $Q^2$.

The investigation of the transition
$\gamma^* N \rightarrow P_{33}(1232)$,
using the experimental data on the pion
photo-and electroproduction on the nucleons,
is connected with the problem of separation
of the resonance and nonresonance contributions
in the multipole amplitudes $M_{1+}^{3/2},E_{1+}^{3/2},S_{1+}^{3/2}$,
which carry information on this transition.
These amplitudes may contain significant nonresonance
contributions, the fact which was clear
with obtaining the first accurate data \cite{3,4} on the 
amplitude $E_{1+}^{3/2}$ at $Q^2~=0$.
The energetic behaviour of this amplitude,
in fact, is incompatible with
the resonance behaviour.
The first investigations of this problem \cite{5,6,7} showed
that it is closely related to the problem
of fulfilment of unitarity condition,
which for electroproduction amplitudes
in the $P_{33}(1232)$
resonance region means the fulfilment
of the Watson theorem\cite{8}:
\begin{equation}
M(W,Q^2)=
\exp(i\delta_{1+}^{\frac{3}{2}}(W))
|M(W,Q^2)|.
\label{1}\end{equation}
Here $M(W,Q^2)$ denotes any of the multipoles under
consideration, and $\delta_{1+}^{\frac{3}{2}}$
is the phase of the corresponding $\pi N$ scattering 
amplitude $h_{1+}^{3/2}(W)=\sin(\delta_{1+}^{3/2}(W))\exp(i\delta_{1+}^{3/2}(W))$.

There are different approaches for the extraction 
of an information on the $\gamma^* N \rightarrow P_{33}(1232)$
transition from the pion photo-and electroproduction data 
with the different forms of the unitarization of the
multipole amplitudes.
These approaches  can be subdivided into the following
groups: the phenomenological approaches \cite{5,6,7,9}
including the approaches based on the K-matrix
formalism \cite{10,11}, the effective Lagrangian approaches
\cite{12,13,14,15,16} with different phenomenological
form of unitarization of amplitudes, 
the dynamical approaches \cite{17,18,19,20,21,22,23},
and the approaches based on the fixed-t
dispersion relations \cite{24,25,26}.

In this work our analysis will be based 
on the solutions for the multipole amplitudes
$M_{1+}^{3/2},E_{1+}^{3/2},S_{1+}^{3/2}$
obtained in Ref. \cite{26}
using the fixed-t dispersion relations
within the approach of Refs.\cite{27,28}.
This approach is very usefull for the 
extraction of an information on 
the $\gamma^* N \rightarrow P_{33}(1232)$
transition, because it in a natural way reproduces
the resonance and noresonance contributions into the
multipole amplitudes, and the obtained solutions 
satisfy unitarity condition (\ref{1}).
Let us discuss this in more detail
using the simplified version of the
dispersion relations
for these multipoles with the s-channel cut only,
i.e. in the form which is similar to
the dispersion relations in the quantum mechanics:
\begin{equation} 
M(W,Q^2)=M^B(W,Q^2)+
\frac{1}{\pi}\int\limits_{W_{thr}}^{\infty}
\frac{ImM(W',Q^2)}{W'-W-i\varepsilon}dW.'
\label{2}\end{equation} 
Here $M^B(W,Q^2)$ is the contribution
of the Born term (i.e. of the nucleon and pion poles)
into the multipoles.
As it was discussed in more detail in Ref.\cite{26},
we can write
in the integrand
of (\ref{2}) 
$ImM(W,Q^2)=h^* (W)M(W,Q^2)$
due to the fact that  
the $\pi N$
ampltude $h_{1+}^{3/2}(W)$ is elastic
up to quite large energies.
Thus,
the dispersion relation (\ref{2}) 
transforms into the singular integral equation
which has a solution in the following analytical form
(see Ref.\cite{27} and the references therein):
\begin{equation}
M(W,Q^2)=M^{part}(W,Q^2)+c_M M^{hom}(W),
\label{3}\end{equation}
where
\begin{equation}
M^{part}(W,Q^2)=M^B(W,Q^2)+
\frac{1}{\pi}\frac{1}{D(W)}
\int\limits_{W_{thr}}^{\infty}
\frac{D(W')h(W')M^B(W',Q^2)}{W'-W-i\varepsilon}dW'
\label{4}\end{equation}
is the particular solution of the singular equation,
generated by the Born term, and
\begin{equation}
M^{hom}(W)=\frac{1}{D(W)}=
\exp\left[\frac{W}{\pi}
\int\limits_{W_{thr}}^{\infty}
\frac{\delta (W')}{W'(W'-W-i\varepsilon)}dW'\right]
\label{5}\end{equation}
is the solution of the homogeneous equation
\begin{equation}
M^{hom}(W)=
\frac{1}{\pi}
\int\limits_{W_{thr}}^{\infty}
\frac{h^* (W')M^{hom}(W')}{W'-W-i\varepsilon}dW',
\label{6}\end{equation}
which enters the solution (\ref{3}) with an arbitrary weight,
i.e. multiplied by an arbitrary constant $c_M$.

The analogy with the quantum mechanics shows that
the solution $M^{part}(W,Q^2)$ 
is the modification of the Born contribution produced
by the $\pi N$ rescattering in
the final state (see Ref.\cite{29}, Chapter 9).
This modification unitarizes the Born contribution
which by itself is real:  
\begin{equation}
M^{part}(W,Q^2)=\exp [i \delta (W)]\left[M^B(W,Q^2)\cos \delta (W)+
e^{a(W)}r(W,Q^2)\right],
\label{7}\end{equation}
where
\begin{equation}
r(W,Q^2)=
\frac{P}{\pi}
\int\limits_{W_{thr}}^{\infty}
\frac{e^{-a(W')}\sin \delta (W')M^B(W',Q^2)}{W'-W}dW',
\label{8}\end{equation}
\begin{equation}
a(W)=
\frac{P}{\pi}
\int\limits_{W_{thr}}^{\infty}
\frac{W \delta (W')}{W'(W'-W)}dW'.
\label{9}\end{equation}

So, $M^{part}(W,Q^2)$ should be considered 
as nonresonance background 
to the resonance contribution. 

It is natural to identify with the resonance
contribution the solution $M^{hom}(W)$, 
because the dispersion relation (\ref{2})
takes the form (\ref{6}), when only 
the $P_{33}(1232)$
resonance contribution  in the s-channel is taken into account.
This solution satisfies the unitarity
condition (\ref{1}) too:

\begin{equation}
M^{hom}(W)=
\frac{1}{D(W)}=\exp [i\delta (W)]
e^{a(W)}.
\label{10}\end{equation}

From Eq. (\ref{7}) it is seen that $M^{part}(W,Q^2)$
has nontrivial energy dependence. The factor at
$\exp [i\delta (W)]$ in $M^{part}(W,Q^2)$
is determined mainly by the first term in the brackets and
changes the sign in the vicinity of the resonance. 
The comparison with the experiment
shows that the amplitude  $E_{1+}^{3/2}$ at $Q^2=0$
is described, in fact,  by $M^{part}(W,Q^2=0)$ \cite{26}.
Hence, this amplitude  
is mainly of nonresonance nature, and its
nontrivial energy dependence is due
to the the final state interaction
in the Born term.

It is important to note that such type nonresonance
contributions exist in all dynamical models 
\cite{17,18,19,20,21,22,23}. 
They are produced by rescattering effects in the pole
terms of these models and have the same type
nontrivial energy dependence as  (\ref{7}).
However, by the magnitudes these contributions are quite different,
because their investigations within the models
contain many model uncertainties coming from
the cutoff procedures, the methods of taking
into account off-shell effects and the methods
of the treatment of the gauge invariance. 
These uncertainties are discussed in detail
in Refs.\cite{30,31}.

It is interesting that in the phenomenological
approaches based on the K-matrix formalism \cite{10,11}
and in the effective Lagrangian approach of Ref. \cite{15},
with the unitarization made by the Noelle method \cite{32}
or using the K-matrix ansatz, the nonresonance contributions
into the multipoles $M_{1+}^{3/2},E_{1+}^{3/2},S_{1+}^{3/2}$
have the same kind energy dependence as (\ref{7}).
In these cases such energy behaviour
of the nonresonance contributions is also connected
with the $\pi N$
interaction in the final state.

In Refs. \cite{24,25} at $Q^2=0$ the 
fixed-t dispersion relations are
used in the same way as in Ref. \cite{26}.
However, the interpretation of the
obtained solutions of the integral equations
is different, although the results for the
whole amplitudes $M_{1+}^{3/2},E_{1+}^{3/2}$
are the same as in \cite{26}. 
In order to extract the $P_{33}(1232)$
resonance contribution in Refs. \cite{24,25}
the method of the Speed Plot analysis is used.
As a result, ignoring the physical nature
of $M^{part}(W)$, the resonance contributions
in these parts of the amplitudes are found.

In Sec. II the multipole amplitudes
which are included into the fitting procedure
in our analysis are listned, and the fitted
parameters are specified.
In Sec. III the results of our analysis
of the TJNAF data at
$Q^2~=$ 2.8 and 4 $(GeV/c)^2$ \cite{1}
and of the DESY data at
$Q^2~=~3.2~(GeV/c)^2$ \cite{2}
are presented. The comparison with theoretical
predictions and with the behaviour of the amplitudes,
which is characteristic of the pQCD asymptotics, is made.

\section{Dispersion relations and
parametrization of multipole amplitudes}
In our analysis we use the fixed-t dispersion 
relations for the Ball 
invariant amplitudes $B_1,B_2,B_3,B'_5,B_6,B_8$ \cite{33},
which for the reaction
$\gamma^*p\rightarrow \pi^0 p$
($B_i^{(\pi^0 p)}=B_i^{(0)}+B_i^{(+)}$)
require no subtraction:
\begin{equation} 
Re B_i^{(\pi^0 p)}(s,t,Q^2)=R_i^{(p)}
\left(\frac{1}{s-m^2}+\frac{\eta_i }{u-m^2}\right)
+\frac{P}{\pi }\int \limits_{s_{thr}}^{\infty}
Im B_i^{(\pi^0 p)}(s',t,Q^2)
\left(\frac{1}{s'-s}+\frac{\eta_i }{s'-u}\right) ds'.
\label{11}\end{equation} 
Here $s=(k+p_1)^2,~u=(k-p_2)^2,~t=(k-q)^2, Q^2=-k^2$,
$k,q,p_1,p_2$ are the 4-momenta
of virtual photon, pion, initial and final protons, respectively,
$\eta_1=\eta_2=\eta_6=1,~\eta_3=\eta'_5=\eta_8=-1$,
$s_{thr}=(m+\mu)^2$, m and $\mu$ are masses
of the nucleon and the pion, and
$R_i^{(p)}$ are the residues in the Born pole terms: 
\begin{eqnarray}
&&R_1^{(\pi^0 p)}=ge(F_1^{(p)}+2mF_2^{(p)}),\nonumber \\
&&R_2^{(\pi^0 p)}=-geF_1^{(p)}(Q^2),\nonumber \\
&&R_3^{(\pi^0 p)}=-\frac{ge}{2}F_1^{(p)}(Q^2), \label{12}\\
&&R_5'^{(\pi^0 p)}=\frac{ge}{2}(\mu -Q^2-t)F_2^{(p)}(Q^2),\nonumber \\
&&R_6^{(\pi^0 p)}=2ge F_2^{(p)}(Q^2),\nonumber \\
&&R_8^{(\pi^0 p)}=geF_2^{(p)}(Q^2),\nonumber
\end{eqnarray}
where in accordance with the existing experimental data we have:
\begin{eqnarray}
&&e^2/4\pi=1/137,~~g^2/4\pi=14.5,\nonumber\\
&&F_1^{(p)}(Q^2)=\left( 1+\frac{g^{(p)}\tau}{1+\tau}\right)G_{dip}(Q^2),\nonumber \\
&&F_2^{(p)}(Q^2)=\frac{g^{(p)}}{2m}\, \frac{G_{dip}(Q^2)}{1+\tau},\label{13}\\
&&G_{dip} (Q^2)=1/(1+Q^2/{0.71~(GeV/c)^2}),\nonumber \\
&&\tau =Q^2/4m^2, ~~ g^{(p)}=1.79.\nonumber
\end{eqnarray}
The imaginary parts of the amplitudes
$B_i^{(\pi^0 p)}(s,t,Q^2)$
we obtain using their  expressions through
the intermediate amplitudes $f_i$
(the corresponing formulas are given in our earlier
work \cite{26}) which have
the following decomposition over multipole amplitudes:
\begin{eqnarray}
&&f_1=\sum\left\{(lM_{l+}+E_{l+})P'_{l+1}(x)+\left[(l+1)M_{l-}+E_{l-} \right]
P'_{l-1}(x)\right\},\nonumber \\
&&f_2=\sum\left[(l+1)M_{l+}+lM_{l-} \right] P'_l(x),\nonumber \\
&&f_3=\sum\left[(E_{l+}-M_{l+})P''_{l+1}(x)+(E_{l-}+M_{l-})P''_{l-1}(x)\right],\label{14}\\
&&f_4=\sum(M_{l+}-E_{l+}-M_{l-}-E_{l-})P''_l(x),\nonumber \\
&&f_5=\sum\left[(l+1)S_{l+}P'_{l+1}(x)-lS_{l-}P'_{l-1}(x)\right],\nonumber \\
&&f_6=\sum\left[lS_{l-}-(l+1)S_{l+}\right] P'_l(x),\nonumber
\end{eqnarray}
where $x=\cos \theta$, $\theta$ is the polar angle
of the pion in the c.m.s.
The relations of the amplitudes $f_i$
to the helicity amplitudes and to
the cross section are also given in \cite{26}.

For the resonance multipole amplitudes
$M_{1+}^{3/2},E_{1+}^{3/2},S_{1+}^{3/2}$
we use as an input the solutions of the integral equations
which follow from the dispersion relations for these
amplitudes. According to these solutions
obtained in Ref.\cite{26} the resonance multipoles
are described as the sums of the particular and 
homogeneous solutions of the integral equations.
The particular solutions
which correspond to the nonresonance contributions
into the multipoles
have the definite magnitudes
fixed by the Born terms.
The homogeneous solutions 
corresponding to the resonance contributions
have the definite
shapes fixed by the homogeneous integral
equations which correspond to the dispersion relations
for $M_{1+}^{3/2},E_{1+}^{3/2},S_{1+}^{3/2}$
with the zero Born terms.
The weights of these solutions are arbitrary and
should be found from the requirement 
of the best description
of the experimental data.
So, the resonance multipoles bring into
our analysis three fitting parameters
which are the weights of the resonance contributions
in the multipoles
$M_{1+}^{3/2},E_{1+}^{3/2},S_{1+}^{3/2}$.

In the $P_{33}(1232)$ resonance region
a significant contribution into $Im f_i$
for the reaction
$\gamma^*p\rightarrow \pi^0 p$
can give also the following combinations of the 
nonresonance multipole amplitudes:
\begin{eqnarray}
&&E_{0+}^{(0)}+\frac{1}{3}E_{0+}^{1/2}+\frac{2}{3}E_{0+}^{3/2},\nonumber \\
&&S_{0+}^{(0)}+\frac{1}{3}S_{0+}^{1/2}+\frac{2}{3}S_{0+}^{3/2},\label{15}\\
&&M_{1-}^{3/2}~and~S_{1-}^{3/2}.\nonumber
\end{eqnarray}
This is connected with the fact
that the $\pi N$ phases corresponding
to these multipoles
are large enough, so, their imaginary parts
can be significant.
In order to take into account these multipoles
in $Im f_i$, we have calculated their
real parts from the Born terms,
then the imaginary parts of the multipoles were found
using for the corresponding $\pi N$ phases
the following analitical formulas:
\begin{eqnarray}
&&\delta_{0+}^{1/2}=\frac{75q}{1+2.5q},\nonumber \\
&&\delta_{0+}^{3/2}=-45q[1+(2.2q)^2], \label{16}\\
&&\delta_{1-}^{3/2}=-(6.9q)^3,\nonumber
\end{eqnarray}
where $q$ is the 3-momentum of the pion in the c.m.s. in the $GeV$ units,
the phases are in the $degree$ units , and all numbers are 
in the $GeV^{-1}$ units.
These formulas describe well experimental data on
the phases 
$\delta_{0+}^{1/2},~\delta_{0+}^{3/2},~\delta_{1-}^{3/2}$
\cite{34,35,36}
up to $E_L=(W^2-m^2)/2m=0.5~GeV$.
At larger energies the smooth cutoff 
for the contributions of (\ref{15}) was made.
We have introduced in our analysis 
four additional fitting parameters
in the form of the coefficients at the
combinations (\ref{15}) found
in the above described way. These parameters were allowed
to vary in the narrow region in the vicinity of 1.

In the description of the data in
the $P_{33}(1232)$ resonance region 
the contributions of the resonances
with higher masses, predominantely from 
the second resonance region,
should be taken into account
in the dispersion integrals.
In the region of $Q^2=3-4~GeV^2$
which we analyse in this work
there is no information on the form factors
of these resonances, except $S_{11}(1535)$.
By this reason we begun our analysis
with the DESY data which cover the 
second resonance region. In this analysis we had 
additional fitting parameters for the 
contributions of the amplitudes
$M_{1-}^{1/2},S_{1-}^{1/2}$
for the $P_{11}(1440)$ resonance,
of the amplitudes
$E_{0+}^{1/2},S_{0+}^{1/2}$
for the $S_{11}(1535)$ resonance,
and of the amplitudes
$E_{2-}^{1/2},M_{2-}^{1/2},S_{2-}^{1/2}$
for the $D_{13}(1520)$ resonance.
The contributions of these amplitudes
were described in the Breit-Wigner form
according to the parametrization of Ref.\cite{37}.
For the multipoles $M_{l+},M_{l-},E_{l+},E_{l-}$
it has the form:
\begin{equation}
M_{B-W}(W,Q^2)=
\frac{M{\bf \Gamma}(W,Q^2)}{M^2-W^2-iM {\bf\Gamma}(W,Q^2)}
\left(\frac{ q_r}{ q}\right)^{l+1}
\left(\frac { k}{ k_r}\right)^{l'}.\label{17}
\end{equation}
For the multipoles  $S_{l+},S_{l-}$
the Breit-Wigner parametrization is:
\begin{equation}
S_{B-W}(W,Q^2)=
\frac{M {\bf\Gamma}(W,Q^2)}{M^2-W^2-iM {\bf\Gamma}(W,Q^2)}
\left(\frac{ q_r}{ q}\right)^{l+1}
\left(\frac { k}{ k_r}\right)^{l'+1}.\label{18}
\end{equation}
Here $l'=l$ for $M_{l+},M_{l-},E_{l+},S_{l+}$,
$l'=l-2$ if $l>1$ for $E_{l-},S_{l-}$, and
$l'=1$ for $S_{1-}$,
$M$ and $\Gamma$ are the masses and the widths of the resonances,
$k_r,q_r$  are the photon and pion 3-momenta in the c.m.s.
at $W=M$, and
\begin{equation}
{\bf\Gamma}(W,Q^2)=\Gamma\left(\frac {q}{q_r}\right)^{2l+1}
\left(\frac{{q}_r^2+X^2}{{q}^2+X^2}\right)^l,\nonumber
\end{equation} 
$X=0.35$. 
So, in the analysis of the DESY data
there are 7 additional fitting parameters which are
the coefficients at (\ref{17},\ref{18})
for the above mentioned multipole amplitudes.
These parameters we consider as effective values
for the description of the second resonance region,
because we did not take into account backgrounds in the
multipole amplitudes in this region
and did not include into our analysis
the resonances from higher resonance regions.
Let us note, however, that the value of the amplitude
$E_{0+}$ for the resonance $S_{11}(1535)$
obtained in this analysis
agrees well with the value known from the
analysis of the $\eta$ electroproduction data.

In the analysis of the TJNAF data,
which do not cover the second resonance region,
we used the results for the multipoles from
this region obtained in the analysis of the DESY data 
with the $Q^2$ evolution
corresponding to the results of Ref. \cite{38}.
Then the small variation of the multipoles was allowed.
\section{Results}
The data used in our analysis
are differential cross sections of $\pi^0$ production
on protons at $Q^2~=$ 2.8 and 4 $(GeV/c)^2$ \cite{1}
and $Q^2~=~3.2~(GeV/c)^2$ \cite{2}.
A total 751 and 867 points which extend over an invariant
mass range $W~=~1.11 - 1.39~GeV/c$
were included in the fit  
at $Q^2~=$ 2.8 and 4 $(GeV/c)^2$, respectively.
At  $Q^2~=~3.2~(GeV/c)^2$ 
we have included in the fit
598 points which extend from $W=1.145~GeV$
to $1.595 GeV$. 
The reduced $\chi^2$ obtained in the analyses
were 1.53, 1.18 and 1.35
at $Q^2~=$ 2.8, 3.2 and 4 $(GeV/c)^2$, respectively.
The obtained results for the
multipole amplitudes
$M_{1+}^{3/2},E_{1+}^{3/2},S_{1+}^{3/2}$
are presented on Fig. 1.
On this figure separately the
resonance and nonresonance contributions
into these amplitudes
are also presented.
It is seen that the nonresonance contributions
play a significant role
in the description of the amplitudes,
especially for  $E_{1+}^{3/2}$ and $S_{1+}^{3/2}$.
In the case of $E_{1+}^{3/2}$
the sum of the resonance and nonresonance contributions
gives the nontrivial energy dependence
of the whole amplitude. At all investigated $Q^2$,
$ImE_{1+}^{3/2}$ changes the sign near the resonance.
So, the energy behaviour of this amplitude,
similar to the behaviour at  $Q^2~=0$,
has by the form nonresonance character.

In the center of the  $P_{33}(1232)$
resonance at $W=m_\Delta$ 
the resonance contributions into
the amplitudes
$M_{1+}^{3/2},E_{1+}^{3/2},S_{1+}^{3/2}$ are:
\begin{eqnarray}
&&ImM_{1+}^{3/2}(res)=0.772\pm 0.031,~0.523\pm 0.021,~0.4\pm 0.016,\nonumber \\
&&ImE_{1+}^{3/2}(res)=0.03\pm 0.022,~0.063\pm 0.017,~0.016\pm 0.012,\label{25}\\
&&ImS_{1+}^{3/2}(res)=-0.038\pm 0.022,~-0.052\pm 0.022,~-0.034\pm 0.008 \nonumber
\end{eqnarray}
at $Q^2~=$ 2.8, 3.2 and 4 $(GeV/c)^2$, respectively.

In Fig. 2 our results for the transverse
form factor $G_T$ of the
$\gamma^* N \rightarrow P_{33}(1232)$
transition are presented in comparison with the data
obtained from inclusive experiments
and partly from exclusive data.
These data are taken from Table 5 of Ref. \cite{39}  by
recalculation for our definition of $G_T$
which is related to the magnetic dipole and electric
quadrupole form factors of Ref.\cite{40} by:
\begin{equation}
[G_T(Q^2)]^2=
(|G^*_M|^2+3|G^*_E|^2)
\bigg(\frac{m_\Delta+m}{2m}\bigg)^2.\label{19}
\end{equation}
At large $Q^2$ our definition of $G_T$
coincides with the Stoler's definition from Ref.\cite{39} :
\begin{equation} 
G_T^2=G_T^2(Stoler)\frac{Q^2}{(m_\Delta-m)^2+Q^2}.\label{20}
\end {equation}
The form factor $G_T$ defined by Eq. (\ref{19})
is more suitable for the description
of low $Q^2$ data.
This form factor is related
to the helicity amplitudes of
the
$\gamma^* N \rightarrow P_{33}(1232)$
transition and to the total cross section of
the reaction $\gamma^* p \rightarrow \pi N$
in the following way:
\begin{equation} 
G_T^2=\frac{1}{4\pi \alpha}\bigg(|A^p_{1/2}|^2+|A^p_{3/2}|^2\bigg)
\frac{2m(m^2_\Delta-m^2)}{(m_\Delta-m)^2+Q^2},\label{21}
\end {equation}
\begin{equation} 
\sigma(\gamma^* p \rightarrow \pi N)=4\pi \alpha G_T^2
\frac{(m_\Delta-m)^2+Q^2}{m_\Delta\Gamma (m^2_\Delta-m^2)}.\label{22}
\end {equation}
It can be expressed through
the multipoles $M_{1+}=(2A_{1+}-3B_{1+})/4$
and $E_{1+}=(2A_{1+}+B_{1+})/4$
using Eq. (\ref{21}) and the relations:
\begin{equation}
A_{1+}^{3/2}=-A^p_{1/2}
\bigg(\frac{3km}{8\Gamma\pi qm_\Delta}\bigg)^{1/2},\label{23}
\end{equation}
\begin{equation}
B_{1+}^{3/2}=A^p_{3/2}
\bigg(\frac{km}{2\Gamma\pi qm_\Delta}\bigg)^{1/2}.\label{24}
\end{equation}

Our results for $G_T$ in Fig. 2 are lower than other
data. This is connected with the fact
that they are obtained by taking into account
only resonance contributions in the amplitude $M_{1+}^{3/2}$
which gives the main contribution into $G_T$.
Our results confirm the whole tendency of  the $G_T$
data to fall more rapidly with increasing $Q^2$
than $1/Q^4$. Let us remind, that
in the pQCD asymptotics  $G_T$ behaves as $1/Q^4$ \cite{41,42,43,44,45}.
So, there is no evidence for the presence of the
pQCD contribution in $G_T$ at $Q^2~<~ 4~GeV^2$.

In Figs. 3,4 our results for the 
ratios $E_{1+}^{3/2}/M_{1+}^{3/2}$
and $S_{1+}^{3/2}/M_{1+}^{3/2}$
corresponding to the resonance contributions to
$M_{1+}^{3/2},E_{1+}^{3/2},S_{1+}^{3/2}$
are presented
together with the data at smaller $Q^2$\cite{46}.
We have presented also
the data points at $Q^2=3.2~GeV^2$
obtained from the DESY data in Ref.\cite{47},
assuming that the multipoles
$M_{1+}^{3/2},E_{1+}^{3/2},S_{1+}^{3/2}$
are described
by the sums of the resonance contributions taken in 
the Breit-Wigner form and the smooth nonresonance backgrounds. 
 
It is known that the information
on the $Q^2$ evolution of
$E_{1+}^{3/2}/M_{1+}^{3/2}$
is important
for the investigation of the  $Q^2$ region 
where the QCD asymptotics begin to work.
This is connected with the fact
that the transition from the quark model prediction
at $Q^2=0$: $E_{1+}^{3/2}/M_{1+}^{3/2}=0$,
to the pQCD asymptotics:
$E_{1+}^{3/2}/M_{1+}^{3/2}\rightarrow 1,~~Q^2\rightarrow \infty$
\cite{41,42,43,44,45},
is characterized by a
striking change of the behaviour of this ratio.
Summarizing our results
one can say that the ratio $E_{1+}^{3/2}/M_{1+}^{3/2}$
is positive
at $Q^2=2.8-4~GeV^2$.
However, by the magnitude it is small,
and the comparison 
with the data at low  $Q^2$ does not
show a visible change in the behaviour
of this ratio with increasing 
$Q^2$. Therefore,
there is no evidence for the presence of the visible
pQCD contribution into the transition
$\gamma^* N \rightarrow P_{33}(1232)$
at $Q^2=2.8-4~GeV^2$.

In Figs. 3,4 the predictions obtained 
in the light cone relativistic
quark model in Refs. \cite{44,45} and
in the relativized
versions of the quark model in Refs.
\cite{48,49}  
are presented. It is seen that the predictions
of \cite{44,45} are in not bad agreement with the data.
We have also presented the predictions from Ref. \cite{50},
where an attempt is made to find some 
approximate formula for the ratio $E_{1+}^{3/2}/M_{1+}^{3/2}$,
which connects the quark model prediction at 
$Q^2=0$ with the pQCD asymptotics.
One of the curves, which corresponds
to a larger asymptotic value of $A_{1/2}$,
describe the  data quite well.

Figures 5-10 are presented to show the typical agreement 
of our results with experimental data.
\section{Discussion}
In this work  
we have analysed  
the TJNAF\cite{1} and DESY\cite{2} data on the 
cross sections of the exclusive reaction
$p(e,e'p)\pi^0$
at
$Q^2~=$ 2.8, 3.2 and 4 $(GeV/c)^2$ 
and found the $P_{33}(1232)$
resonance contribution into the multipole amplitudes
$M_{1+}^{3/2},E_{1+}^{3/2},S_{1+}^{3/2}$.
As an input for the resonance and nonresonance
contributions into these amplitudes
the solutions of the integral equations
for the multipoles obtained
in Ref. \cite{26} were used.
These integral equations follow from the dispersion relations
for $M_{1+}^{3/2},E_{1+}^{3/2},S_{1+}^{3/2}$,
if we take into account the initarity condition 
for the multipoles.
As it was discussed in the Introduction
on the example of the simplified version of the
dispersion relations for the multipoles
with the s-channel cut only,
the solutions of the integral equations
for  $M_{1+}^{3/2},E_{1+}^{3/2},S_{1+}^{3/2}$ contain
two parts which have an interpretation in terms
of the resonance and nonresonace contributions into the
multipoles. One part is the particular solution
of the integral equations generated by the Born term.
This part is the modification of the Born contribution, produced
by the $\pi N$ rescattering in the final state;
we consider it as the nonresonance background contribution.
It has the definite magnitude fixed by the Born term.
Other part of the solutions corresponds
to the homogeneous parts of the integral equations.
We identify it with the resonance contributions.
These solutions have the definite shapes fixed by the dispersion
relations and arbitrary weights which
determine the resonance contributions
into 
$M_{1+}^{3/2},E_{1+}^{3/2},S_{1+}^{3/2}$.
These weights were fitting parameters
in our analyses and were found from the experiment.

The dispersion relations for the multipoles
$M(W,Q^2)\equiv M_{1+}^{3/2},E_{1+}^{3/2},S_{1+}^{3/2}$,
which were investigated in Ref. \cite{26},
in addition to the integrals over s-channel
cut in (\ref{2}) contain also the integrals
over u-channel cut. These integrals, in addition 
to the contribution of $ImM(W',Q^2)$,
include contributions of other multipoles.
The existing information at $Q^2=0$
allows to estimate these contributions, as well as the
high energy contributions into the dispersion integrals.
The calculations made in Ref. \cite{26}
had shown that at $Q^2=0$
all these contributions can be neglected
in comparison with the contribution of the Born term.
So, the particular solutions at $Q^2=0$
are determined by the Born term only.
The information at $Q^2 \ne 0$
is not enough to estimate the contributions
additional to the Born term.
The solutions for  $M^{part}(W,Q^2)$
at $Q^2 \ne 0$ were obtained in  Ref. \cite{26}
under assumption that these solutions
are also determined by the Born term only.
In this work we have used these solutions.
In the future, when experimental data in the 
whole resonance region will be available,
the assumption on the dominance of the Born term contributions
in the terms, which determine the inhomogeneouty
of the integral equations for $M_{1+}^{3/2},E_{1+}^{3/2},S_{1+}^{3/2}$,
will be checked. If it will be found
that the additional contributions
to the Born term are important, a new analysis
in the $P_{33}(1232)$ resonance region,
taking into account these additional contributions, 
will be nessesary.

Let us draw attention to the following point too.
The contributions of the diagram, 
corresponding to the process
$\gamma^* N \rightarrow P_{33}(1232)\rightarrow \pi N $,
into the multipole amplitudes $M_{1+}^{3/2},E_{1+}^{3/2},S_{1+}^{3/2}$
we identify
with the solutions of the homogeneous parts
of the integral equations which follow from the 
dispersion relations for these amplitudes.
The rescattering effects connected with
the $\pi N$ interaction in the final state
modify the $\pi N P_{33}(1232)$ vertex
in this diagram.
A conclusion on the form of this modification
can be made using the results 
of the dynamical model of Ref. \cite{19},
if the amplitude $h_{1+}^{3/2}$
of $\pi N$ scattering is the pure resonance amplitude.
According to these results in this case
the factor at $1/(W-m_\Delta-i\Gamma/2)$ 
for $\gamma^* N \rightarrow P_{33}(1232)\rightarrow \pi N $
is equal to the product
of the vertex $\gamma^* N P_{33}(1232)$
and the dressed vertex $\pi N P_{33}(1232)$.
The dressed vertex $\pi N P_{33}(1232)$
can be found from experimental data on the width
of the $P_{33}(1232)\rightarrow \pi N$ decay.
This fact was used in the derivation
of the relations (\ref{23}),(\ref{24}), which connect
the helicity amplitudes $A^p_{1/2},~A^p_{3/2}$ and
the resonance parts of the amplitudes
$A_{1+}^{3/2},~B_{1+}^{3/2}$
(i.e. of the amplitudes $M_{1+}^{3/2},~E_{1+}^{3/2}$ (\ref{25})).
Our results for the transverse
form factor $G_T$ of the
$\gamma^* N \rightarrow P_{33}(1232)$
transition presented in Fig. 2
are found from Eq. (\ref{21})
using these relations between 
$A^p_{1/2},~A^p_{3/2}$ and $M_{1+}^{3/2},~E_{1+}^{3/2}$ (\ref{25}).

The situation is more complicated,
if the amplitude $h_{1+}^{3/2}$
contains nonresonance background.
In this case it is reasonable to assume,
that the ratios of the resonance parts
of  the multipole amplitudes $M_{1+}^{3/2},E_{1+}^{3/2},S_{1+}^{3/2}$
are equal to the ratios
of the verteces $\gamma^* N P_{33}(1232)$
for these amplitudes, i.e.
the final state interaction modifies 
the $P_{33}(1232)$ resonance
contributions into $M_{1+}^{3/2},E_{1+}^{3/2},S_{1+}^{3/2}$
in the same way.
This assumption is confirmed by the results
obtained in Ref. \cite{21} within
dynamical model.
Therefore, our results
for the ratios  $E_{1+}^{3/2}/M_{1+}^{3/2}$ and
$S_{1+}^{3/2}/M_{1+}^{3/2}$, presented in Figs. 3,4
can be reliably identified with the corresponding
ratios for the $\gamma^* N \rightarrow P_{33}(1232)$
transition.
The same statement is right for
the ratios of the multipole amplitudes at different values 
of $Q^2$, i.e., for example, for the ratios of $G_T$ 
at different values of $Q^2$.
\newline
\begin{center}
{\large {\bf {Acknowledgments}}}
\end{center}

We are  grateful to V. Burkert, B. Mecking and P.Stoler 
for interest and stimulating discussions.
One of us (IGA) has been helped
enormously by interest and insightful questions
of N.Isgur. IGA is also thankful to 
A.V. Radyushkin and N.L.Ter-Isaakyan
for useful discussions. 
The hospitality
at Jefferson Lab, where the main part
of this work was accomplished,
is gratefuly acknowledged by IGA.
SGS thanks the Department of Energy for support
under Contract DE-AC05-84ER40150.

\newpage
{\Large \bf {Figure Captions}}
\vspace{1cm}
\newline
{\large \bf{Fig. 1}} Our results for  
the imaginary parts of the multipole amplitudes
$M_{1+}^{3/2},E_{1+}^{3/2},S_{1+}^{3/2}$.
Dashed curves are the resonance parts of the multipoles
corresponding to the $P_{33}(1232)$ resonance
contribution; dotted curves are the nonresonance background
contributions; full curves are the sums of these contributions;
$E_L=(W^2-m^2)/2m$.
\newline
{\large \bf{Fig. 2}} Experimental data for the transverse
form factor of  the $\gamma N\rightarrow P_{33}(1232)$
transition defined by Eq. (\ref{19}). The data
are divided by $3G_{dip}$, where $G_{dip}(Q^2)=1/(1+Q^2/0.71~ (GeV/c)^2)$.
Data denoted by boxes are taken from Table 5 of Ref. \cite{39}  by
recalculation for our definition of $G_T$; data denoted by asterisks
are obtained in our analysis.
\newline
{\large \bf{Fig. 3}} Experimental data for the ratio
$E_{1+}^{3/2}/M_{1+}^{3/2}$ obtained in our analysis
(asterisks) and the data 
at low $Q^2$ \cite{46} and at $Q^2=3.2(GeV/c)^2$
from  Ref.\cite{47}
in comparison with the predictions 
of Refs.\cite{45} (full line), \cite{48} (dotted line),
\cite{49} (dashed line), \cite{50} (dash-dotted lines).
\newline
{\large \bf{Fig. 4}} Experimental data for the ratio
$S_{1+}^{3/2}/M_{1+}^{3/2}$ obtained in our analysis
(asterisks) and the data 
at low $Q^2$ \cite{46} and at $Q^2=3.2(GeV/c)^2$
from  Ref.\cite{47}
in comparison with the predictions
of Refs.\cite{45} (full line), \cite{48} (dotted line),
\cite{49} (dashed line).
\newline
{\large \bf{Fig. 5}} Comparison of our results for 
$\phi$ distributions with the TJNAF data \cite{1}
at $W=1.235~GeV$ and $Q^2=2.8~(GeV/c)^2$; 
$\epsilon=0.56$.
\newline
{\large \bf{Fig. 6}} Comparison of our results for
energy dependence of the cross sections with the TJNAF data \cite{1}
at $Q^2=2.8~(GeV/c)^2$; $cos\theta=0.7$,
$\epsilon=0.56$.
\newline
{\large \bf{Fig. 7}} Comparison of our results for 
angular distributions with the DESY data \cite{2}
at $W=1.235~GeV$ and $Q^2=3.2~(GeV/c)^2$; 
$\epsilon=0.89$.
\newline
{\large \bf{Fig. 8}} Comparison of our results for
energy dependence of the cross sections with the DESY data \cite{2}
at $Q^2=3.2~(GeV/c)^2$; $\phi=61.5^{\circ}$,
$\epsilon=0.89$.
\newline
{\large \bf{Fig. 9}} Comparison of our results for 
$\phi$ distributions with the TJNAF data \cite{1}
at $W=1.235~GeV$ and $Q^2=4~(GeV/c)^2$; 
$\epsilon=0.51$.
\newline
{\large \bf{Fig. 10}} Comparison of our results for
energy dependence of the cross sections with the TJNAF data \cite{1}
at $Q^2=4~(GeV/c)^2$; $cos\theta=0.7$,
$\epsilon=0.51$.

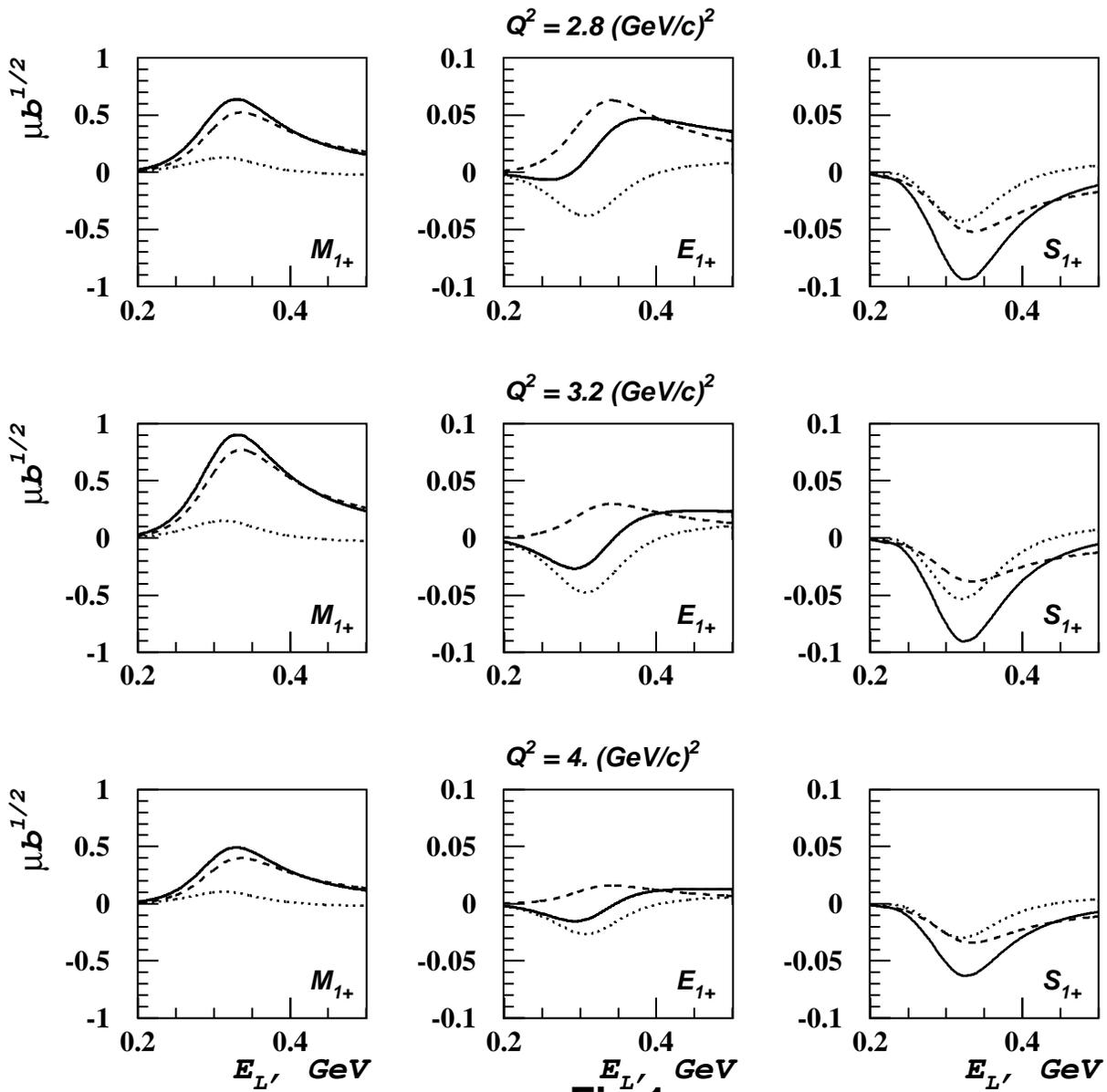
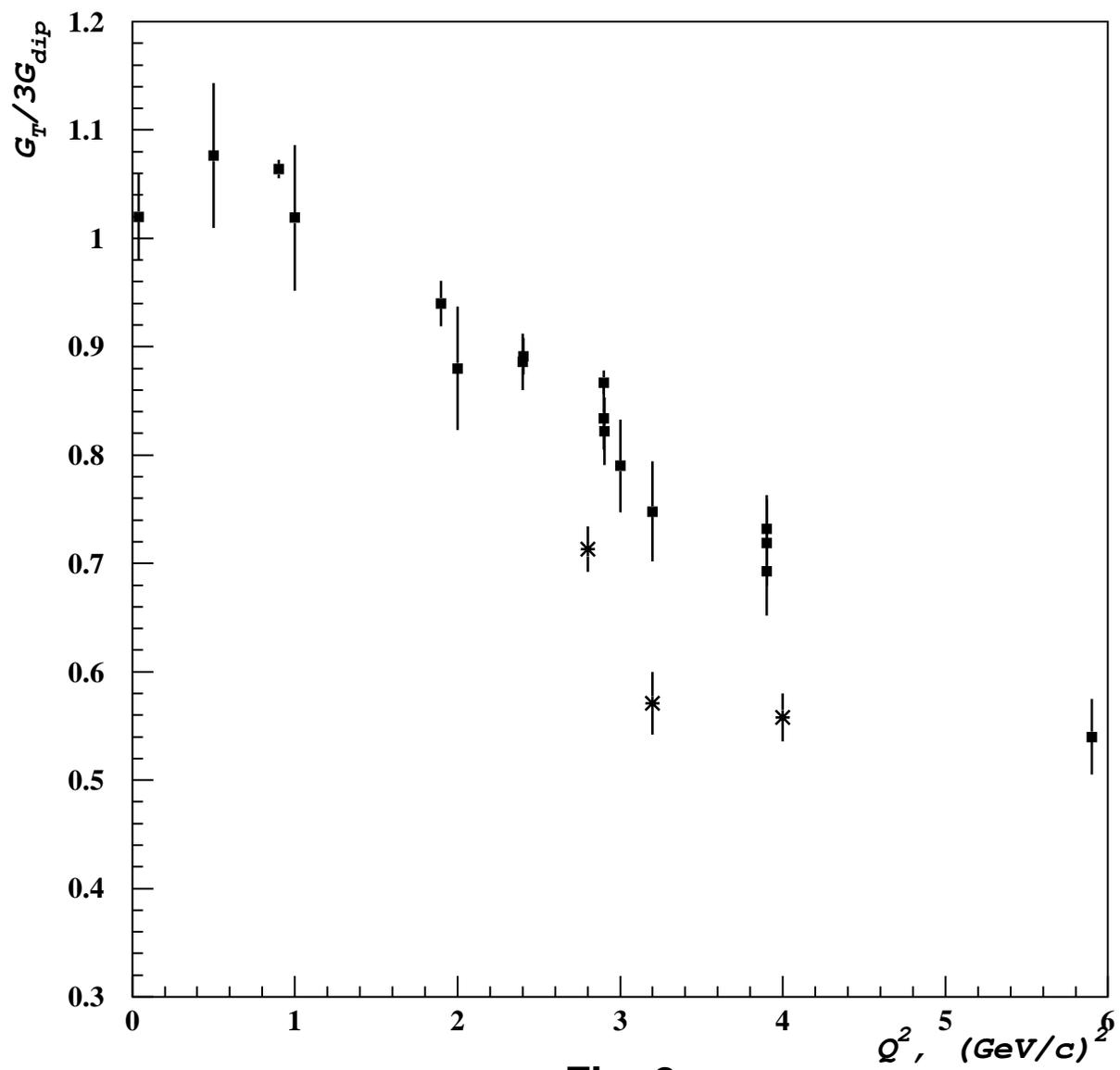
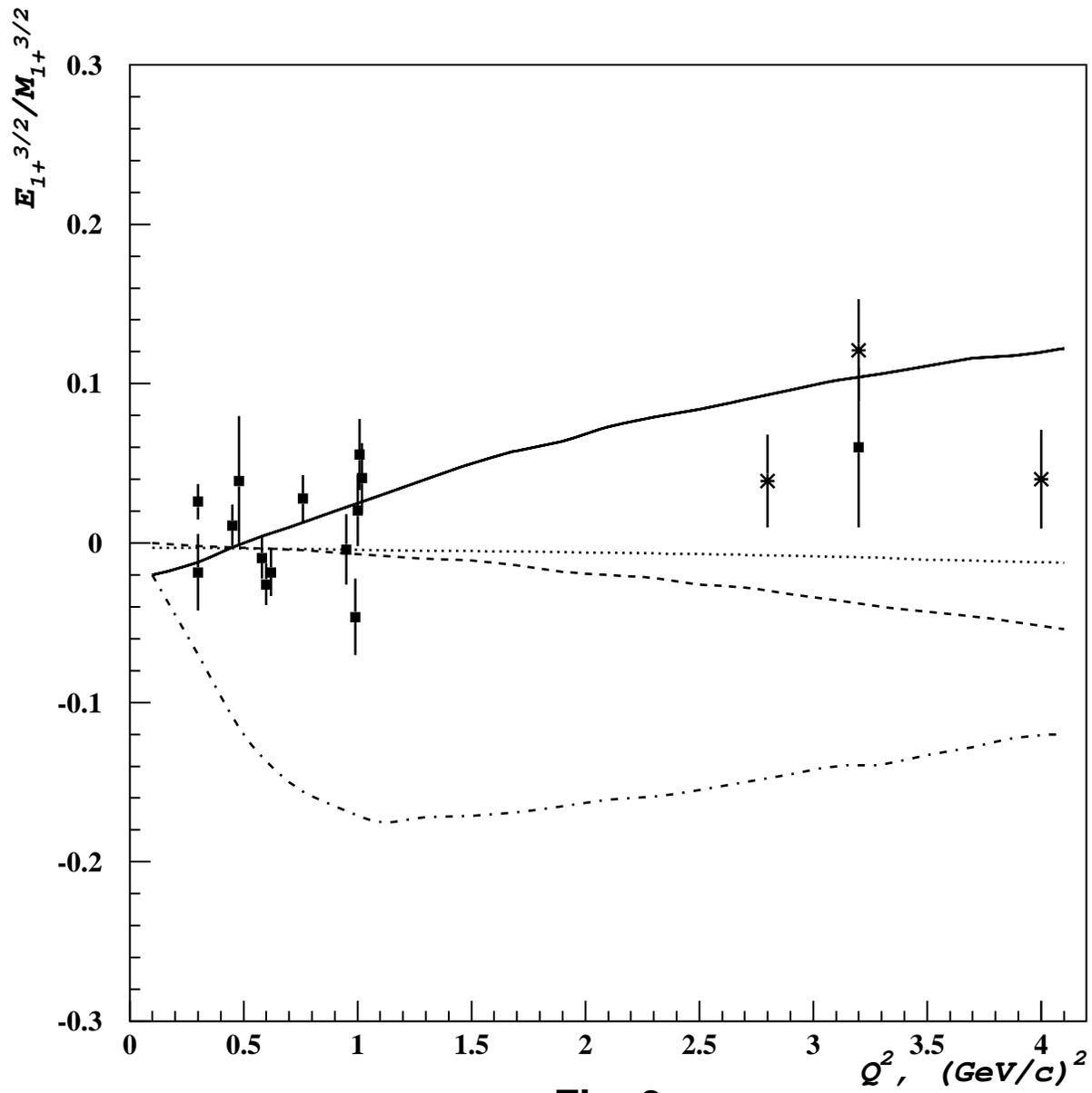
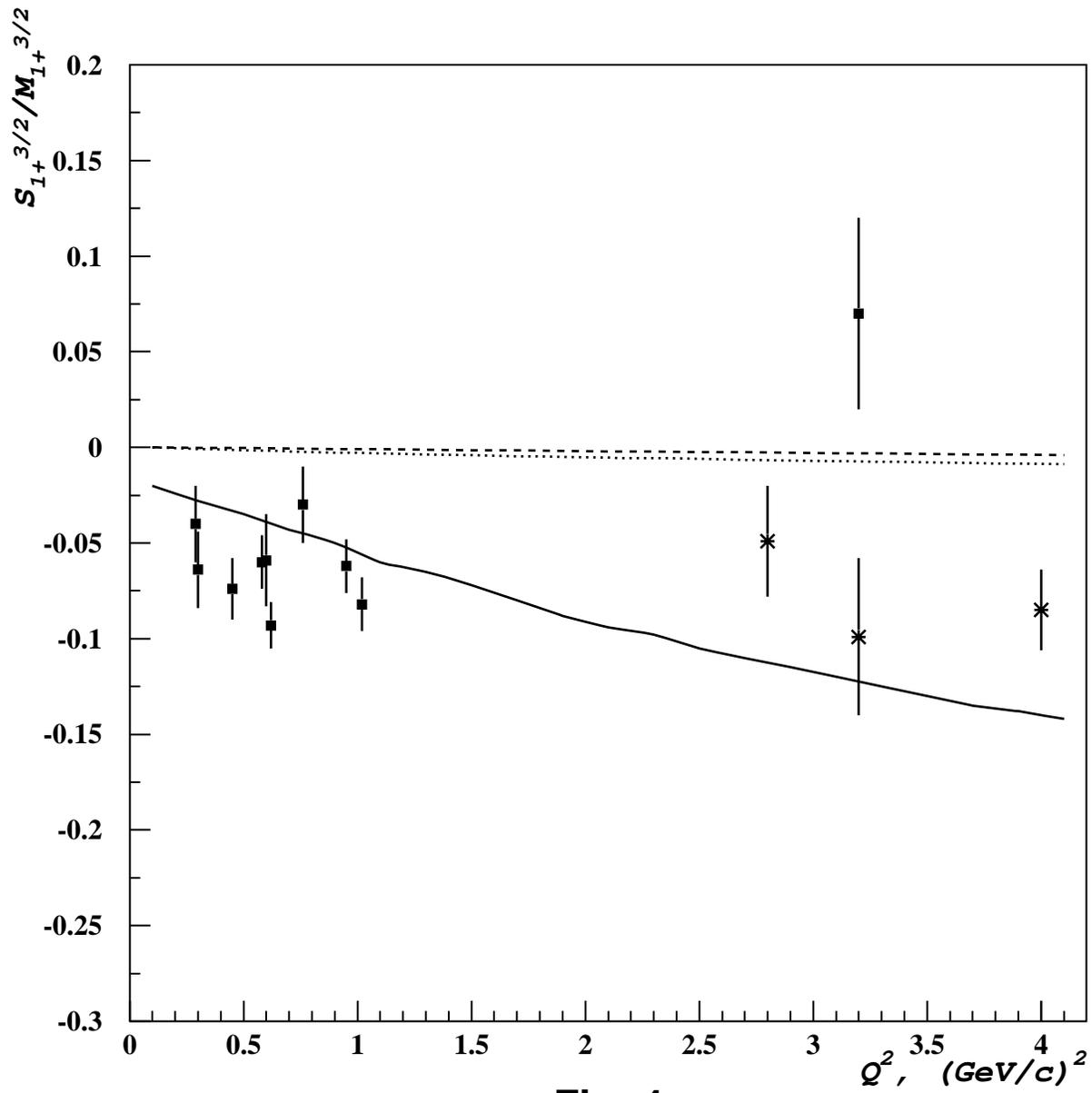
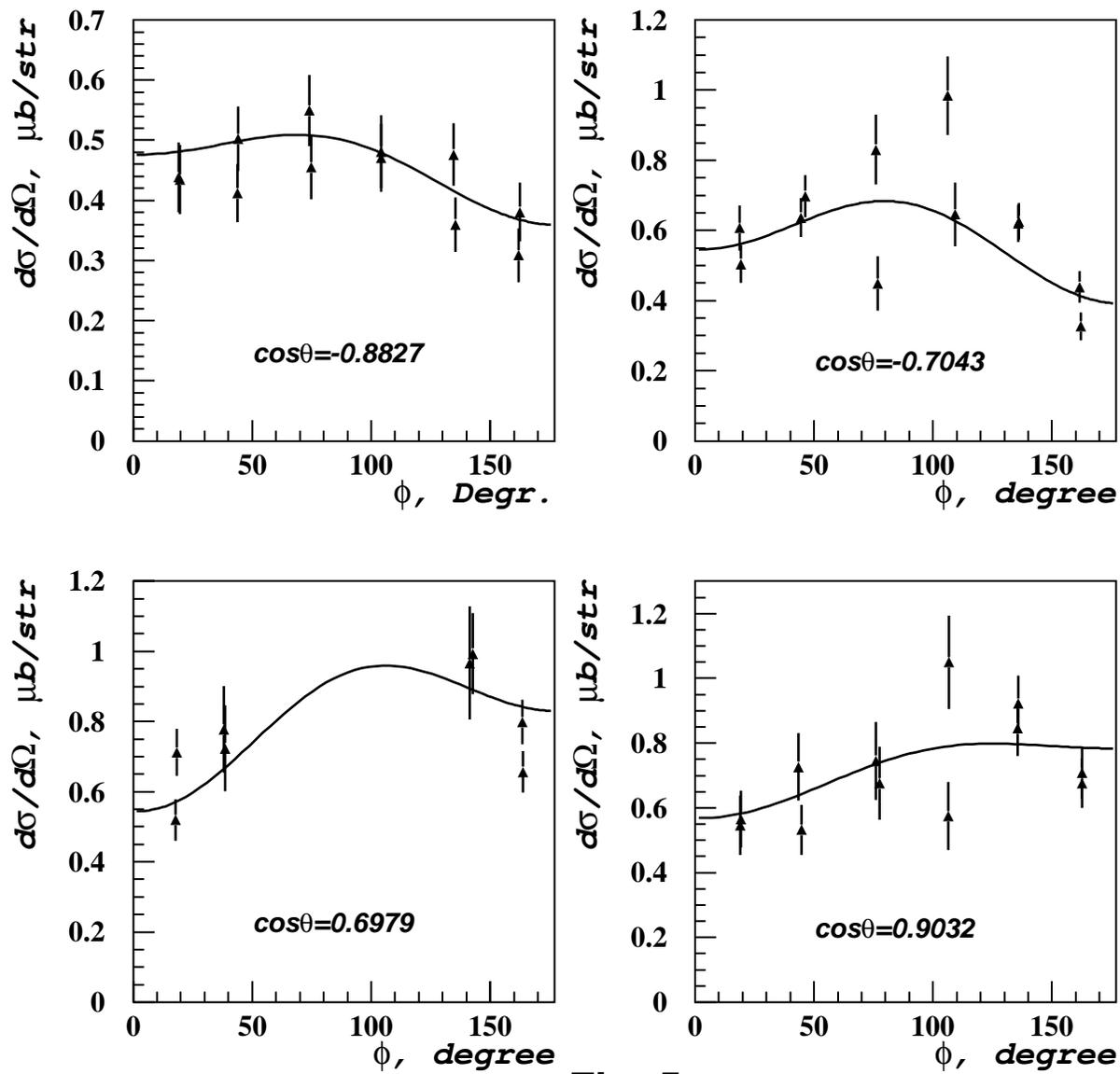
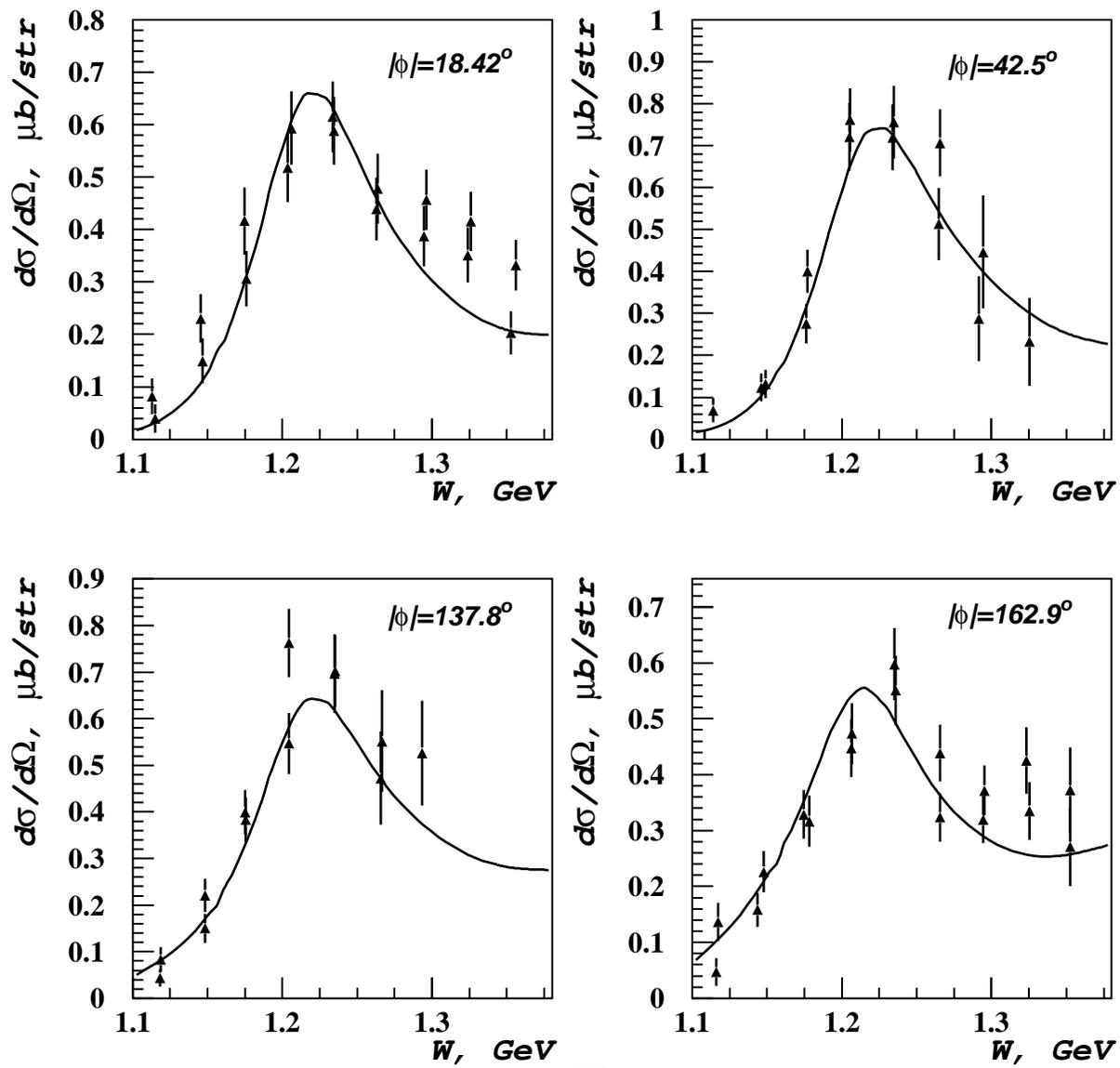
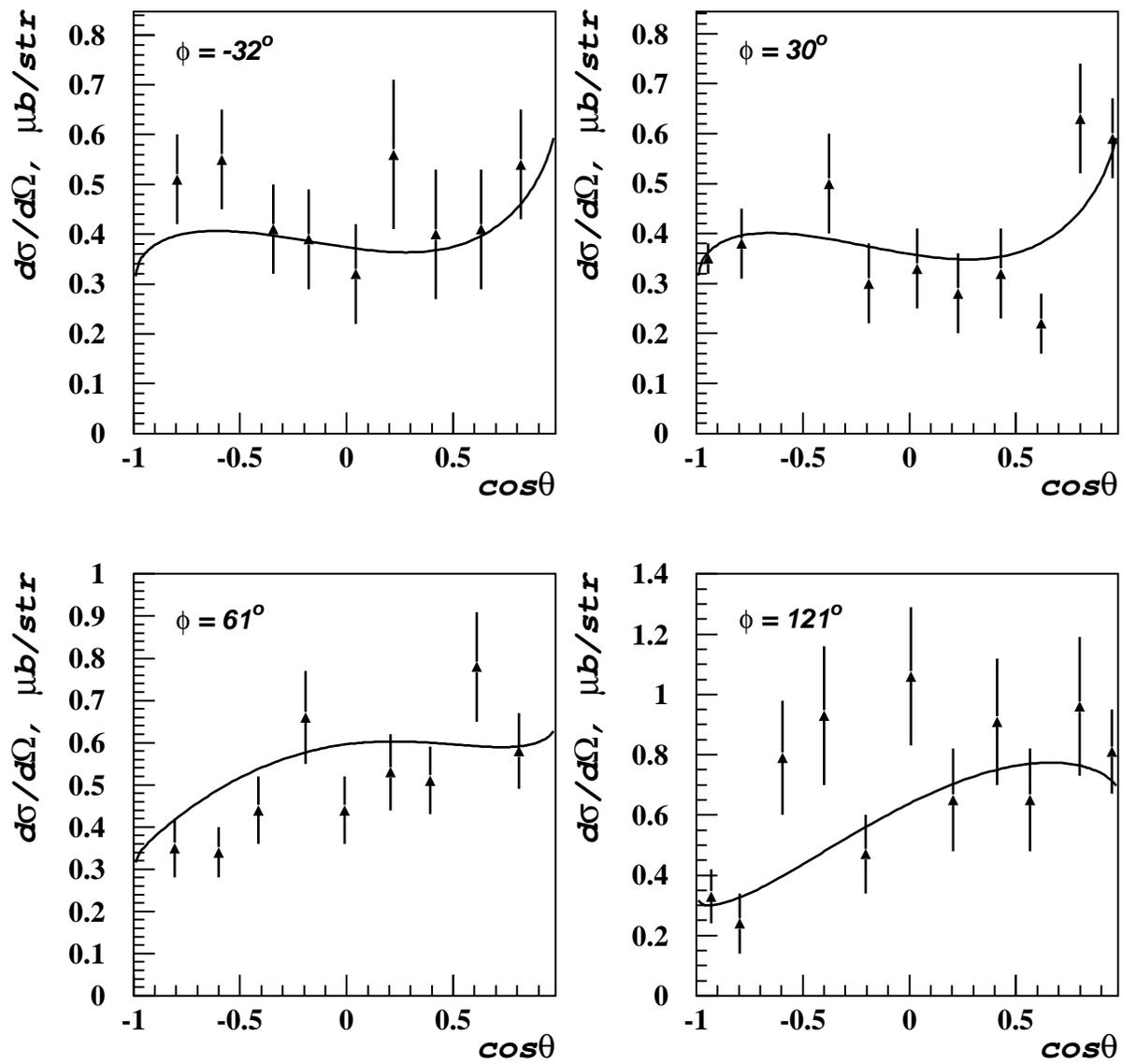
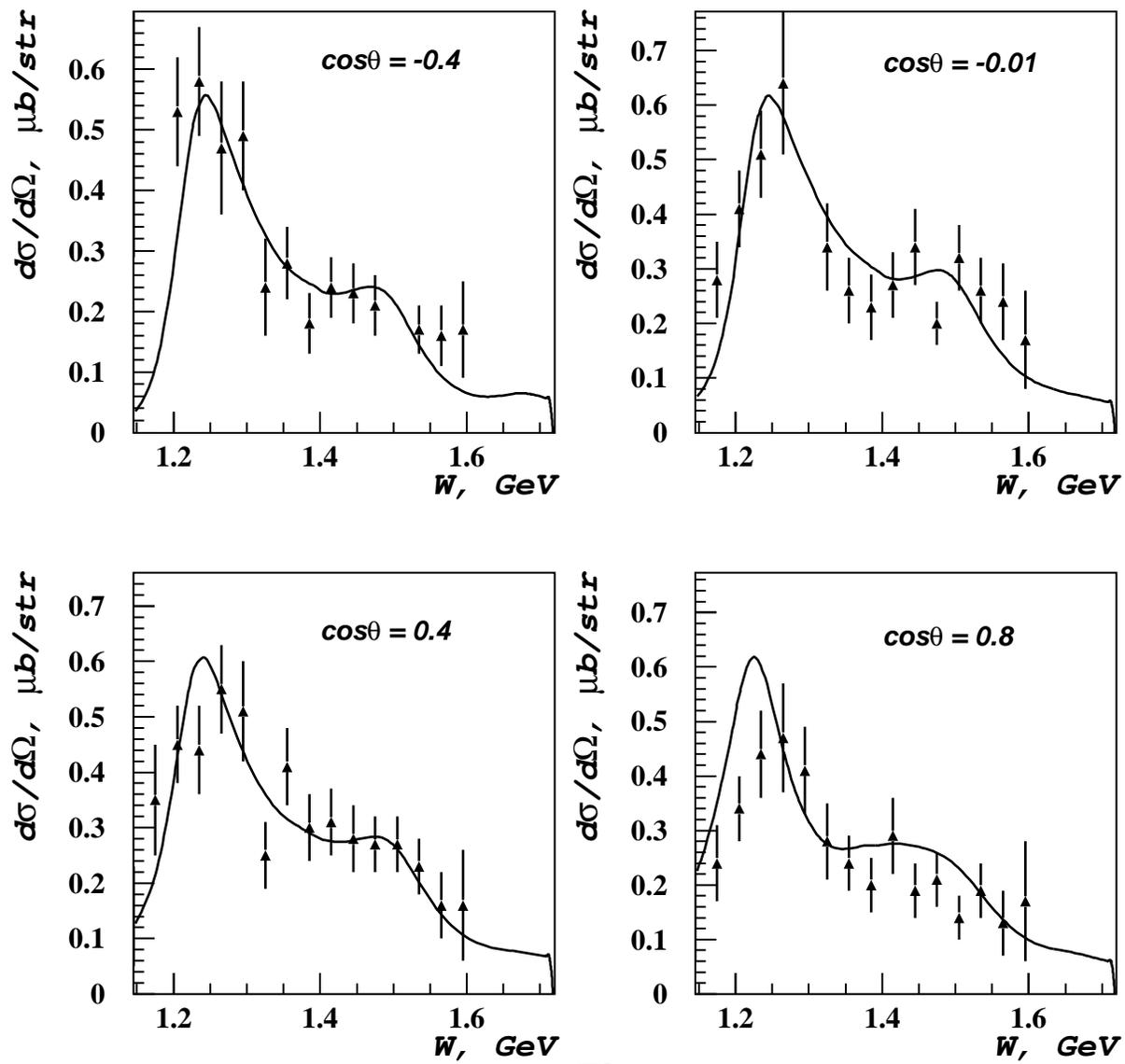
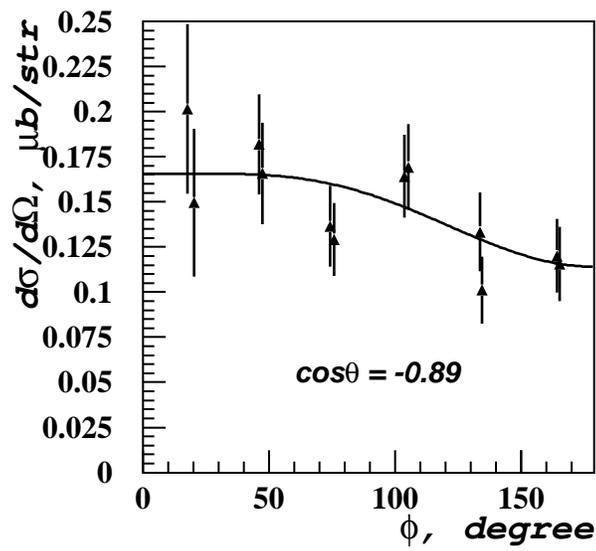
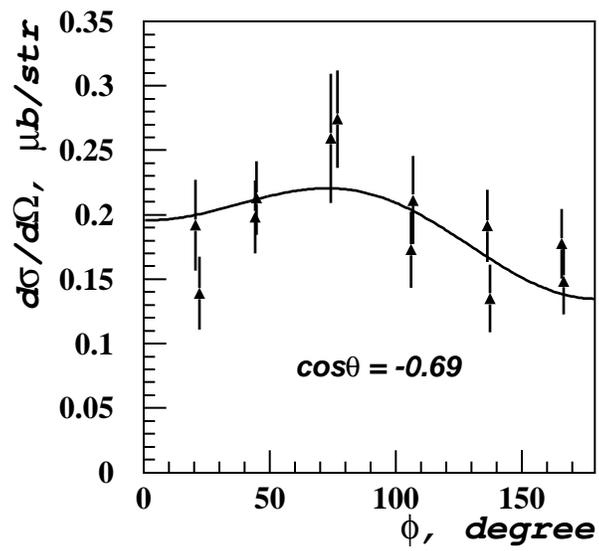
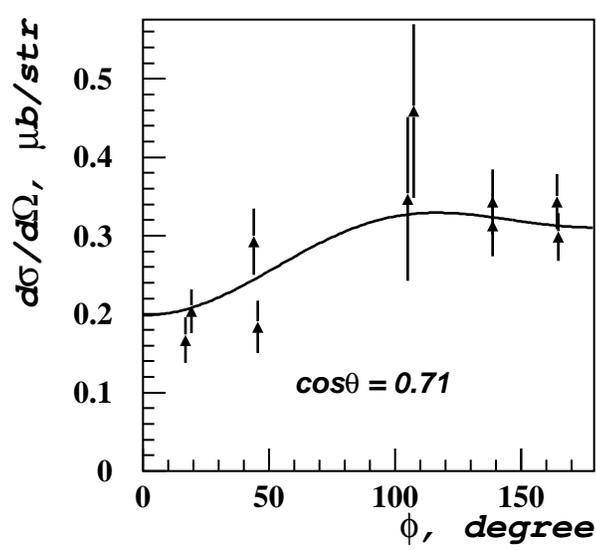
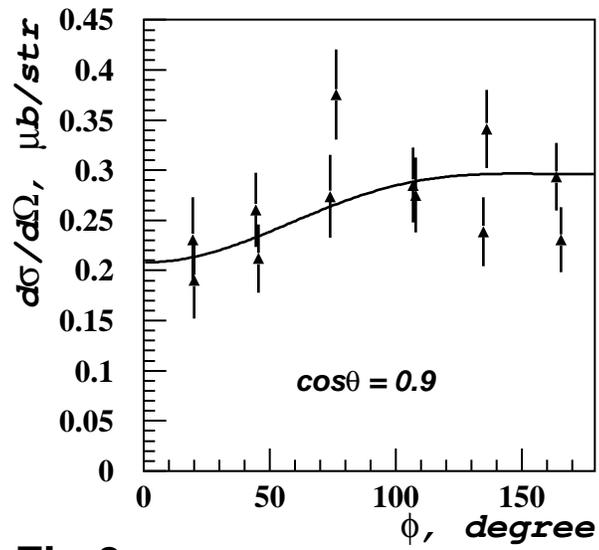
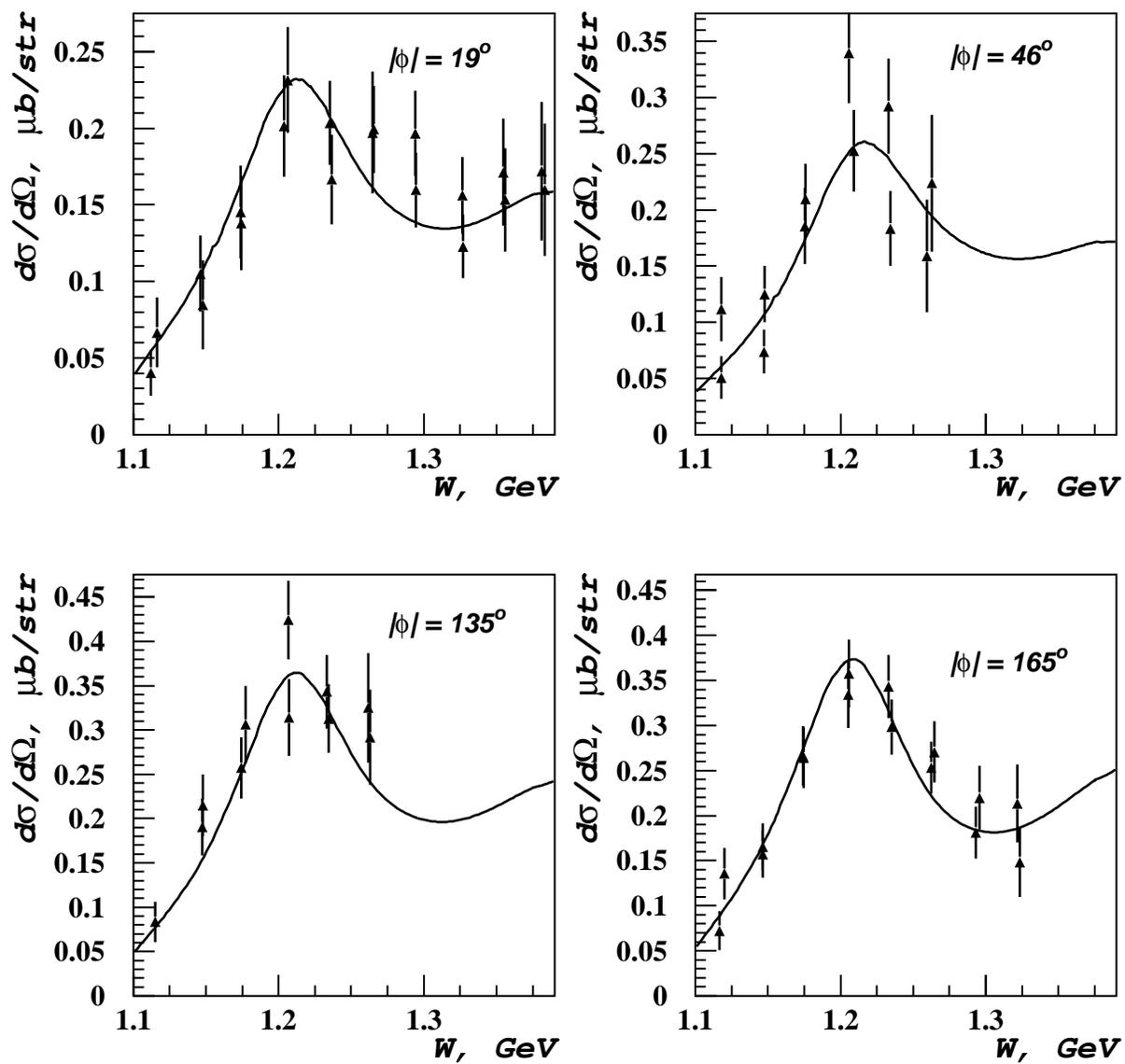
\end{document}